\documentclass[a4paper]{JAC2003}

\usepackage{graphicx}
\usepackage{booktabs}
\usepackage{txfonts}
\usepackage{microtype}
\usepackage{varwidth}
\usepackage{xcolor}

\usepackage[normalsize]{subfigure}
\usepackage{tikz,pgfplots}
\pgfplotsset{compat=newest}
\pgfplotsset{plot coordinates/math parser=false}
\newlength\figureheight
\newlength\figurewidth

\newcommand{\ut}[1]{\,\mathrm{#1}}
\newcommand{\app}{\mathord{\sim}}

\begin{document}
\title{OBSERVATIONS FROM LHC PROTON--PROTON PHYSICS OPERATION}

\author{M. Hostettler,\thanks{michael.hostettler@cern.ch, michihostettler@students.unibe.ch} University of Bern, Switzerland \\G. Papotti, CERN, Geneva, Switzerland}

\maketitle

\begin{abstract}
This paper describes two distinct effects observed during the operation of the LHC in 2012: first, the impacts on beam parameter evolution of the end-of-squeeze instabilities encountered in the second half of the 2012 run; and, second, the very reproducible loss pattern of Beam 1 observed (while a similar pattern was negligible, if present at all, for Beam 2). Statistics for 2012 are provided and the impact on luminosity production is highlighted.
\end{abstract}

\section{Introduction}
\subsection{The LHC Operational Cycle}
The LHC operational cycle for proton physics consists of injecting beams of proton bunches from the injector complex into both rings, accelerating them from injection energy ($450\ut{GeV}$) to the flat-top energy ($4\ut{TeV}$ in 2012), and bringing them into collisions. A full cycle, called a \emph{fill}, is divided into different phases, which are commonly referred to as \emph{beam modes}. This paper covers two effects observed in the part of the cycle after the acceleration phase, which consists of the following beam modes:
\begin{itemize}
\item \emph{Squeeze}: The betatron squeeze, in which the currently separated beams are squeezed to the target collision optics (in 2012: $\beta^*=60\ut{cm}$).
\item \emph{Adjust}: The phase in which the separation between the beams in the interaction points is made to collapse and the beams are brought into collisions.
\item \emph{Stable beams}: The beam mode manually declared by the operators after all adjustments have been made, signalling to the experiments the start of physics data-taking. Physics production fills generally remain in this beam mode for several hours, until the beams are eventually dumped.
\end{itemize}

\subsection{The LHC Filling Scheme}
\begin{figure}[htb]
\setlength\figureheight{20mm}
\setlength\figurewidth{75mm}
%
%
%
%
\begin{tikzpicture}

\begin{axis}[%
small,
width=\figurewidth,
height=\figureheight,
axis on top,
scale only axis,
xmin=0.5, xmax=3564.5,
xtick={1,500,1000,1500,2000,2500,3000,3500},
xticklabels={1,5000,10 000,15 000,20 000,25 000,30 000,35 000},
xlabel={RF Bucket Number ($h=35640$)},
y dir=reverse,
ymin=0.5, ymax=1.5,
ytick={\empty},
axis on top
]
\addplot graphics [xmin=0.5,xmax=3564.5,ymin=0.5,ymax=1.5] {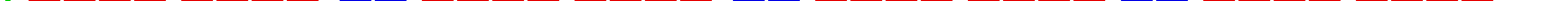};
\end{axis}
\end{tikzpicture}%
   \centering
   \caption{The LHC filling scheme for Beam 1 used in 2012, consisting of eight batches of 144 bunches (red), three batches of 72 bunches (blue), and six witness bunches (green). The filling scheme for Beam 2 is identical apart from the position of the witness bunches.}
   \label{fillingScheme}
\end{figure}

The LHC features a $400\ut{MHz}$ RF system corresponding to a bucket length of $2.5\ut{ns}$ and a harmonic number of $h=35\,640$ \cite{fillingscheme}. For most proton physics production fills in 2012, the LHC was filled with $50\ut{ns}$ spaced bunches. From the SPS, eight batches of 144 bunches, three batches of 72 bunches, and one batch of six bunches (witness bunches for transfer line verification) were injected into each ring of the LHC, totalling 1374 bunches per beam \cite{footnote1}, as shown in Fig.~\ref{fillingScheme}. 

\section{Beam parameter evolution after the end-of-squeeze instabilities}
In the second half of 2012, instabilities were frequently observed at the end of the squeeze beam mode \cite{instability1} in the LHC. Despite not causing significant intensity losses or beam dumps, these instabilities lead to a non-negligible transverse emittance increase of $\app0.5\ut{\mu m}$ for the affected bunches.

\begin{figure}[htb]
\setlength\figureheight{40mm}
\setlength\figurewidth{55mm}
\input{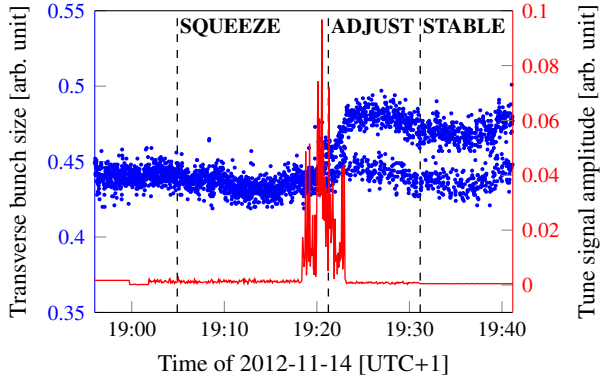}
   \centering
   \caption{The horizontal bunch sizes of Beam 1 (measured for a different bunch every second by the BSRT) and the amplitude of the vertical tune signal (measured by the BBQ system) for fill 3287, indicating the presence of an instability at $\app$19:20. Note the two distinct emittance families thereafter.}
   \label{squeezeInstability}
\end{figure}

Figure~\ref{squeezeInstability} shows, in blue, the horizontal \cite{footnote2} bunch size measurement acquired for a different bunch every second from the scanning Synchrotron Light Telescope (BSRT) system and, in red, the amplitude of the vertical Base-Band-Tune (BBQ) measurement. It can be seen that at $\app$19:20, the BBQ amplitude increases, indicating the presence of an instability. At the same time, certain bunches develop a higher transverse emittance (the horizontal bunch size measurements separate into two bands).

When the beams that experienced the end-of-squeeze instabilities go into stable beams, two distinct families of bunches can be seen, according to the evolution of their parameters: bunches with a larger emittance develop both higher losses and a shorter bunch length, possibly due to limits in the off-momentum dynamic aperture. This effect has historically been labelled `Bunch length histogram splitting', as it was visible on the fixed displays in the LHC control room as a double-peaked histogram of Beam 1 bunch lengths. It has been discussed in detail in a previous publication \cite{instability_michi}, so it will not be treated any further in this paper. The underlying instabilities are subject to current beam--beam studies \cite{instability1}.

The impact on the luminosity is mainly emittance-driven: the emittance, derived from the luminosity, at the start of the stable beams period was generally $\app2.4\ut{\mu m}$ in 2012, while the emittance of bunches affected by the instability was $\app3\ut{\mu m}$, with up to 70\% of all bunches affected for particular fills in late 2012. This corresponds to loss of up to 10\% in both peak and integrated luminosity; for example, from a peak instantaneous luminosity of more than $7000\ut{\mu b^{-1} \cdot s^{-1}}$ for `good' fills to $\app6500\ut{\mu b^{-1} \cdot s^{-1}}$ for fills with $\app50\%$ bunches affected.

\section{The loss pattern of Beam 1 in stable beams}
A very reproducible loss pattern was observed during long physics fills in 2012: the integrated losses of the first $\app30$ bunches of each SPS batch in Beam~1 are up to 10\% lower compared to later bunches after $11\ut{h}$ in stable beams, while such a pattern was always negligible, if present at all, for Beam~2. In the following analysis, only the batches of 144 bunches are considered for simplicity, although the batches of 72 bunches show the same behaviour. A similar pattern had already been noticed during 2011 operation \cite{losspattern_evian}.

\begin{figure}[htb]
\setlength\figureheight{32.5mm}
\setlength\figurewidth{60mm}
\subfigure[The total intensity losses.]{
\input{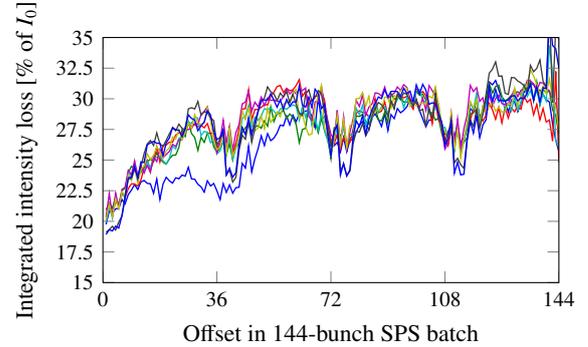}
}
\subfigure[The intensity losses with the burn-off component removed.]{
\input{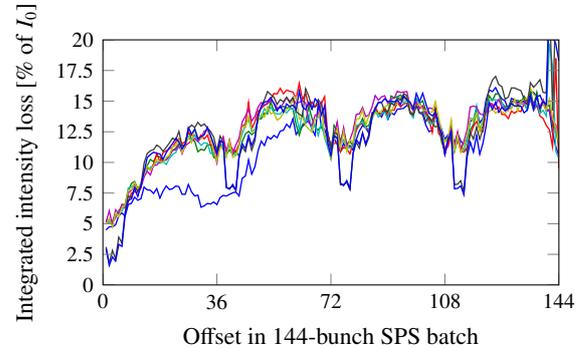}
}
   \centering
   \caption{The integrated losses of Beam 1 for fill 3363 after $11\ut{h}$ in stable beams, grouped by SPS batches of 144 bunches each (a). Note that removing the burn-off component does not change the loss pattern (b).}
   \label{losspatterns}
\end{figure}

The bunch-by-bunch luminosity published by the main experiments and the total process cross-section \cite{cs_totem} allow the intensity lost due to luminosity burn-off to be calculated. Removing the burn-off component from the total losses does not change the overall loss pattern, as depicted in Fig.~\ref{losspatterns}. The cause of the pattern is as yet unknown; in particular, no correlation with the number of long-range interactions has been identified. However, the loss structure of Beam 1 clearly depends on the preceding gaps with no beam (see Fig.~\ref{fillingScheme}): after each 36 bunches, a local decrease in losses is visible after the SPS injection kicker gap, and the first bunches in the first SPS batch after the large LHC dump kicker gap lose even less compared to the same bunches in later SPS batches (the lower blue curves in Fig.~\ref{losspatterns}).

\subsection{Statistics and Correlation to the Bunch Length}
\begin{figure}[htb]
\setlength\figureheight{32.5mm}
\setlength\figurewidth{60mm}

\input{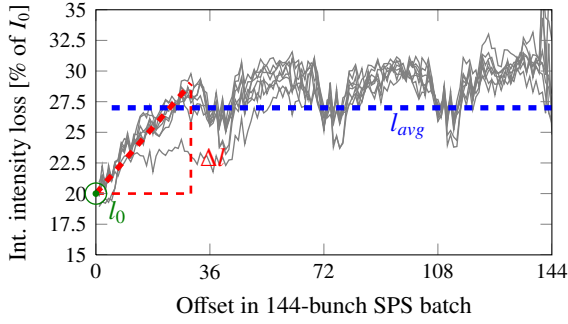}
   \centering
   \caption{Quantitative analysis of the Beam 1 loss pattern for fill 3363, with the average loss of the first bunch of each 144 SPS batch in green, the average slope over the first 30 bunches of each SPS batch in red, and the total average loss in blue.}
   \label{lossstats_expl}
\end{figure}

Fitting a linear function to the integrated losses of the first 30 bunches of each SPS batch allows the difference in losses among those bunches and therefore the strength of the effect to be quantified; averaging over all 144 bunches of each SPS batch shows the impact on the total losses. This is shown in Fig.~\ref{lossstats_expl} for one sample fill.

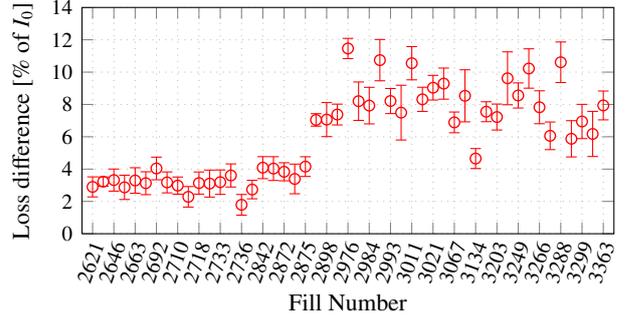
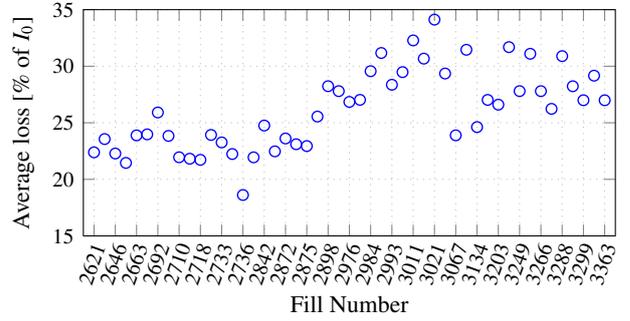
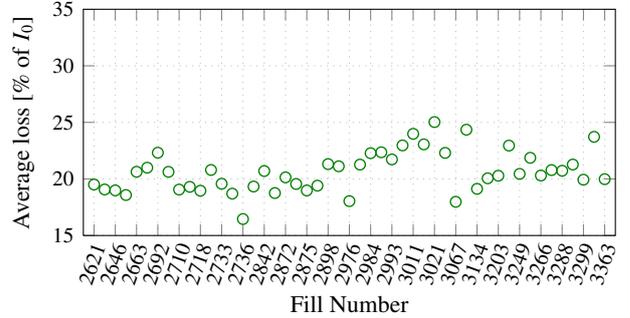
\begin{figure}[htb]
\setlength\figureheight{30mm}
\setlength\figurewidth{70mm}
\subfigure[The difference in the average loss, $\Delta l$, over the first 30 bunches of each SPS batch. The error bars indicate the fit quality.]{
%
%
%
%
\begin{tikzpicture}

\begin{axis}[%
small,
width=\figurewidth,
height=\figureheight,
scale only axis,
xmin=0, xmax=50,
xtick={1,3,5,7,9,11,13,15,17,19,21,23,25,27,29,31,33,35,37,39,41,43,45,47,49},
xticklabels={2621,2646,2663,2692,2710,2718,2733,2736,2842,2872,2875,2898,2976,2984,2993,3011,3021,3067,3134,3203,3249,3266,3288,3299,3363},
xticklabel style={ inner sep=0pt, anchor=north east, rotate=70, xshift=-2pt },
xlabel={Fill Number},
xmajorgrids,
ymin=0, ymax=0.14,
ytick={0,0.02,0.04,0.06,0.08,0.1,0.12,0.14},
yticklabels={0,2,4,6,8,10,12,14},
ylabel={Loss difference [\% of $I_0$]},
ymajorgrids,
grid style={dotted}
]
\addplot [
color=red,
line width=0.5pt,
mark size=2.0pt,
only marks,
mark=o,
mark options={solid},
forget plot
]
plot [error bars/.cd, y dir = both, y explicit]
coordinates{
(1,0.0289764405917455) +- (0.0,0.00619398894316467)(2,0.0322281122967307) +- (0.0,0.00288803326605087)(3,0.0331976844539617) +- (0.0,0.00683049605173578)(4,0.0287028588654809) +- (0.0,0.00750987203080576)(5,0.0329387138725523) +- (0.0,0.00789354802234171)(6,0.0312625475012436) +- (0.0,0.00707988017588603)(7,0.0403865518770892) +- (0.0,0.00700235381090546)(8,0.031760737766628) +- (0.0,0.00647134579522776)(9,0.0297917688092965) +- (0.0,0.00529825349905672)(10,0.0228440591641552) +- (0.0,0.0063468150466011)(11,0.0313563575114851) +- (0.0,0.00684995708947565)(12,0.0310317158434797) +- (0.0,0.00834524036873284)(13,0.0319040457106075) +- (0.0,0.00754867609335194)(14,0.0360566556864442) +- (0.0,0.0071393436674829)(15,0.0178990793207334) +- (0.0,0.00639353917112397)(16,0.0273298263710006) +- (0.0,0.00572107792959949)(17,0.0408870901041448) +- (0.0,0.00681364952011063)(18,0.0403174941182748) +- (0.0,0.00735592027886569)(19,0.0383381699152102) +- (0.0,0.00558622340238581)(20,0.0338954946634826) +- (0.0,0.00917559099533666)(21,0.0415382323493315) +- (0.0,0.00609771409791961)(22,0.0704293087223801) +- (0.0,0.00385853174634673)(23,0.0706340300863208) +- (0.0,0.0105332140968512)(24,0.0737518795140112) +- (0.0,0.0064397200896249)(25,0.11463006882012) +- (0.0,0.00627266333054849)(26,0.0820104299288042) +- (0.0,0.0118876587280034)(27,0.079279232866832) +- (0.0,0.0113043197895539)(28,0.107418732786889) +- (0.0,0.0127834865901821)(29,0.0821113671755607) +- (0.0,0.00776077318040586)(30,0.0749404533834278) +- (0.0,0.0169359351996178)(31,0.105514030966828) +- (0.0,0.0102723879476297)(32,0.0831535420642991) +- (0.0,0.00748441476972497)(33,0.0903687319691077) +- (0.0,0.0076176823172007)(34,0.0928862824327659) +- (0.0,0.00963102666978352)(35,0.0688995181441272) +- (0.0,0.00641010250374413)(36,0.0853429325894417) +- (0.0,0.016104091475198)(37,0.0465269631743233) +- (0.0,0.00618509616674914)(38,0.0755001609986639) +- (0.0,0.00616403508057886)(39,0.0722081339271781) +- (0.0,0.00802905082601861)(40,0.0961805596682681) +- (0.0,0.0163685490171674)(41,0.0855472131293191) +- (0.0,0.0077703867750738)(42,0.102240521675792) +- (0.0,0.0121810202762587)(43,0.0782625313750414) +- (0.0,0.0101886117429063)(44,0.0605820093170476) +- (0.0,0.00855163188326507)(45,0.106104723907468) +- (0.0,0.0125538778642535)(46,0.0587463089495874) +- (0.0,0.0112239997707088)(47,0.0694533364398233) +- (0.0,0.0106064821769689)(48,0.0617370786883715) +- (0.0,0.0139683526314535)(49,0.0793875745583671) +- (0.0,0.0089214592529047)};
\end{axis}
\end{tikzpicture}%
}
\subfigure[The average loss, $l_{avg}$, of all bunches.]{
%
%
%
%
\begin{tikzpicture}

\begin{axis}[%
small,
width=\figurewidth,
height=\figureheight,
scale only axis,
xmin=0, xmax=50,
xtick={1,3,5,7,9,11,13,15,17,19,21,23,25,27,29,31,33,35,37,39,41,43,45,47,49},
xticklabels={2621,2646,2663,2692,2710,2718,2733,2736,2842,2872,2875,2898,2976,2984,2993,3011,3021,3067,3134,3203,3249,3266,3288,3299,3363},
xticklabel style={ inner sep=0pt, anchor=north east, rotate=70, xshift=-2pt },
xlabel={Fill Number},
xmajorgrids,
ymin=0.15, ymax=0.35,
ytick={0.15,0.2,0.25,0.3,0.35},
yticklabels={15,20,25,30,35},
ylabel={Average loss [\% of $I_0$]},
ymajorgrids,
grid style={dotted}
]
\addplot [
color=blue,
line width=0.5pt,
mark size=2.0pt,
only marks,
mark=o,
mark options={solid},
forget plot
]
table{
1 0.223882671715596
2 0.235610171013274
3 0.222832521149459
4 0.214527550453972
5 0.238870404052359
6 0.239741937868862
7 0.259110280928846
8 0.238394960305082
9 0.219521003674113
10 0.218220180394086
11 0.217241880607534
12 0.239255190071633
13 0.232627983132717
14 0.22238222078453
15 0.186178291887679
16 0.219449054079334
17 0.247554667749077
18 0.224733267279972
19 0.236123224895386
20 0.231030651764471
21 0.229420554873104
22 0.255501581096133
23 0.282341284020317
24 0.277978319545256
25 0.2684612568828
26 0.270267451126533
27 0.295605593834758
28 0.311703218658456
29 0.283602741510473
30 0.29479403924301
31 0.322793625610097
32 0.306695378510328
33 0.341200346285939
34 0.293544345388894
35 0.238902653876086
36 0.314520224528191
37 0.246207318987801
38 0.270213354932088
39 0.265999499567412
40 0.316882262590956
41 0.277977334883105
42 0.310988375975076
43 0.277951499740316
44 0.262309040332176
45 0.30891554538193
46 0.28229361132843
47 0.269937122966223
48 0.29165216476514
49 0.269917512096189
};
\end{axis}
\end{tikzpicture}%
}
\subfigure[The average loss, $l_0$, of the first bunch of each SPS batch.]{
%
%
%
%
\begin{tikzpicture}

\begin{axis}[%
small,
width=\figurewidth,
height=\figureheight,
scale only axis,
xmin=0, xmax=50,
xtick={1,3,5,7,9,11,13,15,17,19,21,23,25,27,29,31,33,35,37,39,41,43,45,47,49},
xticklabels={2621,2646,2663,2692,2710,2718,2733,2736,2842,2872,2875,2898,2976,2984,2993,3011,3021,3067,3134,3203,3249,3266,3288,3299,3363},
xticklabel style={ inner sep=0pt, anchor=north east, rotate=70, xshift=-2pt },
xlabel={Fill Number},
xmajorgrids,
ymin=0.15, ymax=0.35,
ytick={0.15,0.2,0.25,0.3,0.35},
yticklabels={15,20,25,30,35},
ylabel={Average loss [\% of $I_0$]},
ymajorgrids,
grid style={dotted}
]
\addplot [
color=green!50!black,
line width=0.5pt,
mark size=2.0pt,
only marks,
mark=o,
mark options={solid},
forget plot
]
table{
1 0.195060156045474
2 0.190690776662815
3 0.189885417670006
4 0.185785348774846
5 0.206322049963249
6 0.209961505173482
7 0.223197131136675
8 0.206280028424336
9 0.190601439738287
10 0.193010703023941
11 0.189575733603439
12 0.207937956584129
13 0.195744568738447
14 0.187025400903849
15 0.164563384589663
16 0.193313324895372
17 0.207009067529036
18 0.187528838482681
19 0.201400115411637
20 0.195562447043054
21 0.189823485087711
22 0.194071262288772
23 0.213115693867314
24 0.211203086199592
25 0.180399615183062
26 0.212684314565308
27 0.222761535309722
28 0.223568689994531
29 0.217251170936705
30 0.229632856950258
31 0.239859618409383
32 0.230552578088722
33 0.250278891813651
34 0.223036880954451
35 0.179807115542882
36 0.24351010779664
37 0.191247406894999
38 0.200565796473548
39 0.20285065417837
40 0.229477077127011
41 0.204455727139865
42 0.218684365615747
43 0.203031802929289
44 0.207839723218244
45 0.207232162422205
46 0.212742369671744
47 0.19932676068825
48 0.237216757579567
49 0.19979882345413
};
\end{axis}
\end{tikzpicture}%
}

   \centering
   \caption{The statistics for the loss pattern of Beam 1 after $11\ut{h}$ in stable beams for 2012. Note the sudden increase of both the slope and the average losses after fill 2875, probably related to an increased bunch length.}
   \label{lossstats}
\end{figure}

Figure~\ref{lossstats} shows this analysis applied for all 2012 fills that lasted for at least $11\ut{h}$ in stable beams. It is to be noted that $\Delta l$, the observed difference over the first 30 bunches, suddenly increases from less than 5\% to up to $10\%$ after fill 2875 (see Fig.~\ref{lossstats}(a)). This is suspected to be correlated with the increase of the bunch length target for the ramp from $1.2\ut{ns}$ to $1.3\ut{ns}$ from fill 2880 onwards, indicating that the pattern is correlated with longitudinal losses. An increase of $l_{avg}$, the average intensity loss of all bunches, is also observed for the same fills (Fig.~\ref{lossstats}(b)), while $l_0$, the intensity loss of the first bunch of each SPS batch, remains at the same level (Fig.~\ref{lossstats}(c)).

\subsection{Impact on Luminosity}
Increased losses lead to a lower total intensity and therefore to a lower integrated luminosity for the same fill duration due to a decreased luminosity lifetime, which is defined as $\tau$ in

\begin{equation}\label{llt}
L(t) = L_0 \exp\left(-\frac{t}{\tau}\right).
\end{equation}

\begin{figure}[htb]
\setlength\figureheight{35mm}
\setlength\figurewidth{65mm}
\input{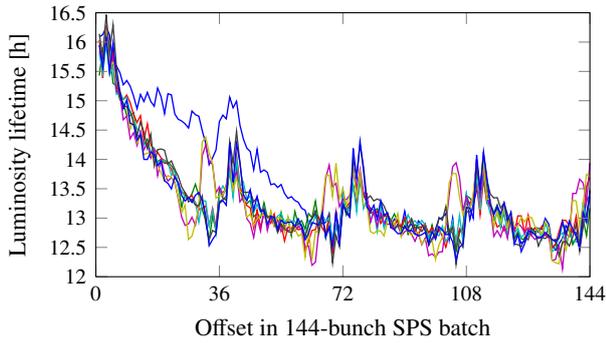}
   \centering
   \caption{The bunch-by-bunch luminosity lifetime for fill 3363, calculated from ATLAS data after $8\ut{h}$ of stable beams. It shows an inverted image of the Beam 1 loss pattern that was observed.}
   \label{losspattern_lumi_lt}
\end{figure}

In the bunch-by-bunch luminosity lifetime calculated by fitting Eq.~\ref{llt} to the luminosity per colliding bunch pair published by the ATLAS experiment for a time window of $2\ut{h}$ set at $8\ut{h}$ after the start of stable beams, the impact of the Beam 1 loss pattern is visible, as depicted in Fig.~\ref{losspattern_lumi_lt}.

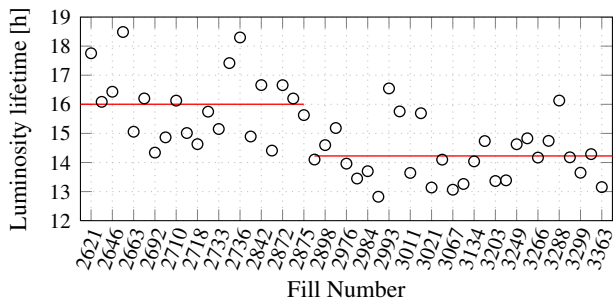
\begin{figure}[htb]
\setlength\figureheight{27mm}
\setlength\figurewidth{70mm}
%
%
%
%
\begin{tikzpicture}

\begin{axis}[%
small,
width=\figurewidth,
height=\figureheight,
scale only axis,
xmin=0, xmax=50,
xmin=0, xmax=50,
xtick={1,3,5,7,9,11,13,15,17,19,21,23,25,27,29,31,33,35,37,39,41,43,45,47,49},
xticklabels={2621,2646,2663,2692,2710,2718,2733,2736,2842,2872,2875,2898,2976,2984,2993,3011,3021,3067,3134,3203,3249,3266,3288,3299,3363},
xticklabel style={ inner sep=0pt, anchor=north east, rotate=70, xshift=-2pt },
xlabel={Fill Number},
xmajorgrids,
ymin=12, ymax=19,
ytick={12,13,14,15,16,17,18,19},
yticklabels={12,13,14,15,16,17,18,19},
ylabel={Luminosity lifetime [h]},
ymajorgrids,
grid style={dotted}
]
\addplot [
color=black,
line width=0.5pt,
mark size=2.0pt,
only marks,
mark=o,
mark options={solid},
forget plot
]
table{
1 17.7548032876114
2 16.0844973784751
3 16.4318637707193
4 18.4856029017235
5 15.057389166355
6 16.201606938097
7 14.3391309901559
8 14.864015620985
9 16.1265971435305
10 15.0123729635432
11 14.6306022515738
12 15.7477203718992
13 15.1511815725167
14 17.4168210187717
15 18.2980175432672
16 14.8937144515095
17 16.6616280693615
18 14.4089580502959
19 16.6584659355516
20 16.1997744439247
21 15.6292023491239
22 14.1012822709484
23 14.5974242756947
24 15.1849210409883
25 13.9609959093885
26 13.4488508392931
27 13.7003876770619
28 12.8225806569846
29 16.5460918358307
30 15.7555281447251
31 13.6414084414404
32 15.6933502348009
33 13.1396785491495
34 14.0976660706335
35 13.0639517080623
36 13.2609963824107
37 14.0382159110454
38 14.7363516488143
39 13.3627233534353
40 13.3865708143374
41 14.6277848517247
42 14.8302849061662
43 14.1688687617778
44 14.7428602413049
45 16.1293298615176
46 14.1774680503032
47 13.6478578580721
48 14.2875547679296
49 13.1541291435011
};
\addplot [
color=red,
line width=0.5pt,
forget plot
]
table{
0 16.002569819952
21 16.002569819952
};
\addplot [
color=red,
line width=0.5pt,
forget plot
]
table{
22 14.2251826502622
50 14.2251826502622
};
\end{axis}
\end{tikzpicture}%
   \centering
   \caption{The total luminosity lifetime, calculated using the ATLAS luminosity data after $8\ut{h}$ in stable beams, for the fills considered in Fig.~\ref{lossstats}. The luminosity lifetime got worse after fill 2875, probably due to the increased total intensity loss. The red lines indicate the average up to and after fill 2875.}
   \label{lossstats_lumi_lt}
\end{figure}

The increased total losses after fill 2875 are also visible on the total luminosity lifetime. As shown in Fig.~\ref{lossstats_lumi_lt}, the luminosity lifetime after fill 2875 is on average $\app1.5\ut{h}$ lower in the long term compared to earlier fills. Despite the rather large spread of the values, the bunch-by-bunch pattern observed (Fig.~\ref{losspattern_lumi_lt}) indicates a correlation with the loss pattern in question. Assuming the same peak luminosity, this decrease in luminosity lifetime corresponds to a $\app7\%$ loss in integrated luminosity for a fill lasting $11\ut{h}$ in stable beams.

\section{Conclusions}
Observations on two distinct effects that affect the LHC luminosity production were studied and presented in this paper. First, instabilities at the end of the betatron squeeze increased the transverse emittances of selected bunches by $\app0.5\ut{\mu m}$. Up to 70\% of all bunches in Beam 1 were affected in late 2012, resulting in a loss of up to 10\% of both peak and integrated luminosity.

Second, bunch-by-bunch observations on the integrated intensity losses of Beam 1 bunches showed a very reproducible pattern building up over several hours in stable beams on 144-bunch SPS batches, while no such pattern was observed on Beam 2. A clear cause for the pattern has not been identified yet, but the differences in losses over a SPS batch, as well as the average losses, increased after an increase in bunch length, indicating a correlation with longitudinal losses. The shorter luminosity lifetime due to the increased losses leads to a loss of $\app7\%$ of integrated luminosity for long fills.

\section{Acknowledgements}
The authors would like to thank G.~Arduini, X.~Buffat, W.~Herr, T.~Pieloni, the LHC Beam-Beam Team, and the LHC Beam Operation Committee for fruitful discussions on the topics covered by this paper.


\begin{thebibliography}{9}
\bibitem{footnote2}
No vertical bunch size measurement was available due to technical difficulties.
\end{thebibliography}
\end{document}